\documentclass[12pt]{article}
\usepackage{amsmath}
\begin{document}
\begin{center}
\LARGE
\textbf{Bohmian trajectories and the ether:
Where does the analogy fail?}\\[1cm]
\large
\textbf{Louis Marchildon}\\[0.5cm]
\normalsize
D\'{e}partement de physique,
Universit\'{e} du Qu\'{e}bec,\\
Trois-Rivi\`{e}res, Qc.\ Canada G9A 5H7\\
email: marchild$\hspace{0.3em}a\hspace{-0.8em}
\bigcirc$uqtr.ca\\
\end{center}
\medskip
\begin{abstract}
Once considered essential to the explanation
of electromagnetic phenomena, the ether was
eventually discarded after the advent of special
relativity.  The lack of empirical signature of
realist interpretative schemes of quantum mechanics,
like Bohmian trajectories, has led some to conclude
that, just like the ether, they can be dispensed
with, replaced by the corresponding emergence of the
concept of information.  Although devices like
Bohmian trajectories
and the ether do present important analogies, I argue
that there is also a crucial difference, related to
distinct explanatory functions of quantum mechanics.
\end{abstract}
\section{Introduction}
Quantum information theory has been one of the
most active areas of development of quantum
mechanics in the past two decades.  The
realization that transfer protocols based
on quantum entanglement may be absolutely
secure has opened new windows in the field of
cryptography (Bennett \& Brassard, 1984).
And the development
of quantum algorithms thought to be exponentially
faster than their best classical counterparts has
drawn great interest in the construction of quantum
computers (Shor, 1994).
These face up extraordinary
challenges on the experimental
side (Vandersypen \emph{et al.}, 2001).
But attempts to build them are likely to throw
much light on the fundamental process of
decoherence (Zurek, 1991)
and perhaps on the limits of quantum mechanics
itself ('t Hooft, 1999; Leggett, 2002).

Along with quantum information theory came also
a reemphasis of the view that the wave function
(or state vector, or density matrix) properly
represents knowledge, or
information (Rovelli, 1996;
Fuchs \& Peres, 2000; Fuchs, 2002).
This is often called the
\emph{epistemic view} of quantum states.  On what
the wave function is knowledge of, proponents of
the epistemic view do not necessarily agree.
The variant most relevant to the present discussion is
that rather than referring to objective
properties of microscopic objects (such as
electrons, photons, etc.), the wave function
encapsulates probabilities of results
of eventual macroscopic measurements.  The Hilbert
space formalism is taken as complete, and its
objects in no need of a realistic interpretation.
Additional constructs, like value
assignments (van Fraassen, 1991; Vermaas, 1999),
multiple worlds (Everett, 1957; DeWitt, 1970;
Wallace, 2003),
or Bohmian trajectories (Bohm, 1952;
Bohm \& Hiley, 1993; Holland, 1993)
are viewed as superfluous at best.

The methodological rule calling to discard
additional constructs to the Hilbert space
has been likened to the one that led to
abandon the concept of the ether in the
early part of the twentieth
century\footnote{Of course
the chronology of added constructs is reversed
in the two episodes.  Ether theories predated
special relativity, whereas Bohmian trajectories
came after the Hilbert space formalism.
Cushing (1998) has argued that history
could plausibly have been reversed in the
latter episode.} (Bub, 2004, 2005;
Bohr, Mottelson, \& Ulfbeck, 2004).
H.~A.~Lorentz and his contemporaries
viewed electromagnetic phenomena as taking place
in a hypothetical medium called the ether.  From
this, Lorentz developed a description of
electromagnetism in moving reference frames,
and he found that the motion is
undetectable (Lorentz, 1909).
Following Einstein's formulation of
the electrodynamics of moving
bodies (Einstein, 1905),
the ether was recognized as
playing no role, and was henceforth discarded.
So should it be, according to most proponents of the
epistemic view of quantum states, with interpretations
of quantum mechanics which posit observer-independent
elements of reality.
They predict no empirical differences
with the Hilbert space formalism, and therefore
should be discarded.

The purpose of this paper is to analyse,
in their respective contexts, the explanatory
roles of the ether and additional constructs to the
Hilbert space formalism.  To be specific, and because
earlier discussions have largely focussed on them,
I shall formulate my argument in terms of Bohmian
trajectories, without however implying any fundamental
commitment to that choice.\footnote{See Tumulka (2004)
and Brown and Wallace (2005) for two different recent
assessments of Bohmian mechanics.}
I will first recall that Bohmian
trajectories coexist rather
well with the notion of a preferred reference frame,
which the ether traditionally defines.
I will next point out
what was involved in the transition between ether
theories and special relativity.  The function of
Bohmian trajectories will then be investigated, in
connection with two distinct explanatory roles of
quantum mechanics.  This will evince a crucial
difference between Bohmian trajectories and the
ether, and illustrate why interpretative schemes
cannot be dispensed with in quantum mechanics.
%
%\newpage
\section{Quantum mechanics and special relativity}
The ether has long been viewed as defining
a preferred inertial frame of reference.
In this sense, it can find room in the conceptual
structure of quantum mechanics, at least in some
of the ways the formalism is presented.

Take, for instance, the theory's highly
influential articulation proposed by
von Neumann (1932).
There the wave function of a quantum system
evolves in two very different ways.  Outside
the context of measurements, it obeys the
Schr\"{o}dinger equation or, equivalently,
it evolves through the action of a unitary
operator (von Neumann's process 2).  At the
end of a measurement, however, it stochastically
transforms into one of the eigenfunctions of the
observable being measured (process 1, or
collapse).\footnote{Von Neumann really writes that
the pure state density matrix transforms into a
mixture, but for a given quantum system, only one
component of the mixture obtains.}  That
evolution is not unitary, and does not obey the
Schr\"{o}dinger equation.

It is not difficult (for one particle at least)
to make process~2 consistent
with the special theory of relativity.  One just has
to replace the Schr\"{o}dinger equation by the Dirac
equation.  Process~1, however, is much more tricky.
The wave function of a quantum system usually
covers the whole of three-dimensional space, that is,
its support is unbounded.  Since the collapse is
taken to occur instantaneously (or very nearly so),
the wave function, as a consequence of measurement,
changes values everywhere at the same time.  It is
very difficult to make this process relativistically
covariant.  Indeed wave function collapse seems to
single out a preferred inertial reference frame.

Due to the statistical character of
the predictions of quantum
mechanics, wave function collapse does not, as
is well known, allow the transfer of information
faster than the speed of light.  In this sense, at
least, it is consistent with special relativity.
It would thus seem that the frame being singled
out cannot be determined experimentally.

Similar considerations can be made in the
context of Bohmian mechanics.  Consider a set of
$N$ particles of masses $m_a$ and charges $e_a$,
in an electromagnetic field specified by the
four-potential $A^{\mu}$.  The Dirac wave function
$\Psi$ then has $4^N$ components, which if needed can
be specified by $N$-tuples of four-valued indices.
Let $\gamma_{a \mu}$ represent Dirac matrices acting
on the $a^{\text{th}}$ index, and let $\alpha_{a k} =
\gamma_{a 0} \gamma_{a k}$.  The Dirac equation
can then be written as
\begin{align}
- \frac{i}{c} \frac{\partial \Psi}{\partial t}
&+ \sum_a \left\{ \frac{e_a}{\hbar} 
A^{0}(\mathbf{r}_a, t) \Psi
- i \alpha_{a k} 
\frac{\partial \Psi}{\partial x_{a k}} \right. \notag\\
& \qquad\qquad \mbox{} + \left. \frac{e_a}{\hbar} \alpha_{a k} 
A^{k}(\mathbf{r}_a, t) \Psi
+ \frac{m_a c}{\hbar} \gamma_{a 0} \Psi \right\} = 0 .
\label{dirac}\end{align}
Eq.~(\ref{dirac}) entails that
\begin{equation}
\frac{\partial j_0}{\partial t}
+ c \sum_a \frac{\partial j_{a k}}{\partial x_{a k}} = 0 ,
\label{conser}\end{equation}
where
\begin{equation}
j_0 = \Psi^{\dagger} \Psi , \qquad
j_{a k} = \Psi^{\dagger} \alpha_{a k} \Psi .
\label{current}\end{equation}
Bohmian trajectories can be introduced by
specifying that the three-velocity of particle $a$ at
the space-time point $(\mathbf{r}_a, t)$ is given by
\begin{equation}
v_a^i = c \, j_a^i (j^0)^{-1} .
\label{velocity}\end{equation}
It can be shown (Bohm \& Hiley, 1993; Holland, 1993)
that the magnitude of the velocity never exceeds $c$,
and that if particles are distributed according
to the probability density $\Psi^{\dagger} \Psi$
at a given time, they will be distributed according
to $\Psi^{\dagger} \Psi$ at any other time.

The $N$-particle Dirac equation being relativistically
covariant (Bohm \& Hiley, 1993),
it can be used indifferently in any inertial
reference frame.  Probabilities obtained in one
frame transform correctly into probabilities
obtained in any other frame.  Since, however,
the quantities $(j^0 , j_a^i)$ do not make up a
four-vector (for $N>1$), the Bohmian trajectories
are not covariant.  That is,
trajectories computed in frame $\Sigma$ through
Eqs.~(\ref{velocity}), when transformed into
$\Sigma '$ by means of the Lorentz transformations,
will not match trajectories computed in $\Sigma '$
directly through~(\ref{velocity}).  Hence
Eqs.~(\ref{velocity}) can hold in only one inertial
reference frame (modulo rotations and space-time
translations).

The upshot is that Bohmian trajectories, like
von Neumann's collapse, naturally coexist with the
ether construed as defining a preferred, albeit
unobservable, reference frame.\footnote{Maudlin
(1994) gives a detailed argument why the introduction
of a preferred frame may be the more rational choice
to make in a quantum-mechanical theory.}
There remains to see
whether the justification of the former, in quantum
mechanics, is of the same nature as that of the latter,
in electromagnetic theories.
%
%\newpage
\section{Ether and field}
The concept of ether has a long
history (Whittaker, 1951; Darrigol, 2000),
and it was used
in a number of different contexts.  In one of these,
mainly developed in the nineteenth century,
the ether was viewed as a substratum wherein
electric and magnetic phenomena take place.
Much effort was spent on detailed mechanical
models of the substratum.  It is not the purpose of this
paper to recapitulate them, but it may be worthwhile
to recall one proposed by Maxwell, as he described it
in an 1861 letter to W.~Thomson (quoted in
Whittaker, 1951, p.~250).
\small
\begin{quote}
I suppose that the ``magnetic medium'' is divided
into small portions or cells, the divisions or cell walls
being composed of a single stratum of spherical
particles, these particles being ``electricity.''
The substance of the cells I suppose to be highly
elastic, both with respect to compression and
distorsion; and I suppose the connection between
the cells and the particles in the cell walls to be such
that there is perfect rolling without slipping between
them and that they act on each other tangentially.

I then find that if the cells are set in rotation,
the medium exerts a stress equivalent to a hydrostatic
pressure combined with a longitudinal tension along
the lines of axes of rotation.
\end{quote}
\normalsize
Maxwell goes on drawing detailed analogies between
his cells and cell walls and ``a system of magnets,
electric currents and bodies capable of magnetic
induction.''  It should be pointed out, however,
that such models played a less important role
in Maxwell's great treatise (1873),
which relied more on a Lagrangian formulation.
Model building was progressively abandoned
in the most fruitful late nineteenth century
contributions to electromagnetic theory, those
of Lorentz in particular.

\begin{sloppypar}
Lorentz's largely definitive views on
electromagnetic theory were expounded in his
1906 Columbia University lectures, published
a few years later.
His ontology is threefold: there is ponderable
matter,\footnote{Lorentz does not commit himself
on whether all matter is made of electric
charges, nor on whether all mass has an
electromagnetic origin.}
there are electric charges (``electrons''),
and there is the ether, ``the receptacle of
electromagnetic energy and the vehicle for many
and perhaps for all the forces acting on
ponderable matter'' (Lorentz, 1909, p.~30).
The ether is supposed to be
at rest, and this determines an absolute inertial
reference frame.
\end{sloppypar}

As there was no reason to expect that the earth is
at rest with respect to the ether,\footnote{Everyone
realized that if the earth is at
rest in the ether now, it should not be in a few
months, due to its orbital velocity around the sun.
Unless perhaps the ether is dragged along the motion,
but this causes numerous other problems.}
the question arose as to how to describe
electromagnetic phenomena in a moving frame.
The calculation would go roughly as follows.
Maxwell's equations, assumed to hold in the frame
where the ether is at rest, would be used to compute
the electric and magnetic fields of moving electrons
and matter.  These fields would act back on charges
and matter, and (partly at least) determine their
configuration.  It could be shown, in particular,
that if all forces in moving matter shared the
characteristics of electric and magnetic forces, the
so-called Lorentz-FitzGerald contraction would
naturally follow.\footnote{Brown (2001) points out
that a simple contraction is not the only deformation
that can account for the null result of the
Michelson-Morley experiment, a fact which Lorentz,
and probably FitzGerald, fully realized.}

In this context, Lorentz and others were able to
prove a rather remarkable result.  Suppose we introduce
spatial coordinates $\mathbf{r}'$ at rest in the
moving frame, and define a ``local time'' $t'$ as
$t' = t - (\mathbf{r}' \cdot \mathbf{w})/c^2$, where
$\mathbf{w}$ is the velocity of the moving frame.
Furthermore, introduce new electric and magnetic
fields related to the old ones by what is now called
a Lorentz transformation.\footnote{The argument
was first made with terms of first order in
$w/c$, and then more generally.}  Then the numerical
values of the new fields coincide with values of the
(old) fields that would be obtained from electrons
and matter having a similar configuration
in the frame where
the ether is at rest.  Moreover, the new fields and
space-time coordinates satisfy Maxwell's equations.

The above result was interpreted, by Lorentz and
most generally by Poincar\'{e}, as showing that motion
with respect to the ether is
undetectable (Poincar\'{e}, 1905; Paty, 1993).
Yet Lorentz and Poincar\'{e} never abandoned the
ether, and Lorentz maintained that the local time is
just a definition, the true time always referring
to the rest frame of the ether.\footnote{Poincar\'{e}
(1898) saw the conventional character of the
simultaneity of distant events.  Later he recognized
that clocks synchronized by means of light signals
mark the local time, which however he contrasted
with the true time (Pais, 1982, Chap.~6).}
It was Einstein's fundamental contribution to view
the local time as the time genuinely measured in
the moving frame, with no more and no less reality
than the time measured in any other inertial frame.
Suddenly the ether was seen as playing no useful role,
and was eventually discarded.

Many now believe that Lorentz's conception of
(unobservable) true time and absolute rest, and
Einstein's notion of complete equivalence between
inertial frames, are both logically consistent and
in agreement with empirical
results (Gr\"{u}nbaum, 1973; Bell, 1976).
Yet in less than a decade, most people adopted
Einstein's views (Pais, 1982).  The great simplicity
of Einstein's purely kinematical approach and the
fact that it allowed complete freedom in the choice
of the inertial frame where calculations would be
made no doubt contributed to that decision.

It is important to realize, however, that the rejection
of the ether has not left a void in its stead.  From
Maxwell's quotation to Lorentz's final views, we have
seen that the ether was progressively deprived of much
of its complicated mechanical attributes.  There only
remained something to define a preferred frame and
transmit the electric and magnetic forces.  But
as the ether was discarded, the electromagnetic field
acquired by itself an independent reality.  For
Einstein, this way of seeing the field was one of the
most important consequences of the conceptual
development leading to special
relativity (Einstein, 1949; Paty, 1993).

Even before the full development of quantum mechanics
and quantum field theory, the electromagnetic
field was generally considered as
being real, as anything carrying energy and momentum
is.  The ether was discarded in its role as defining
absolute time and absolute motion.  The
methodological choice, however, was not one between
the ether and nothing, but one between the ether as
sustaining the field and a self-sustaining field.
%
%\newpage
\section{Bohmian trajectories and information}
Bohmian trajectories and the ether are elements
of two different theoretical structures.  They
present both analogies and differences.  The
analogy that is relevant here is that neither Bohmian
trajectories in quantum mechanics nor the ether in
special relativity lead to specific empirical
consequences.  Does that mean that the trajectories,
or other interpretative devices,
have in quantum mechanics the same status as the
ether in special relativity?  And if one can
dispense with such devices, is there
something which, like the field, plays the
role they would otherwise have?

To examine these questions, it is appropriate
to start with the following observation.
Although all measurements are made
by means of macroscopic apparatus, quantum
mechanics is used, as an explanatory theory,
in two different ways: it is meant to explain
(i) nonclassical correlations between macroscopic
objects and [ultimately through quantum field
theory] (ii) the small-scale structure of
macroscopic objects.  That these two functions
are distinct is shown by considering a hypothetical
situation where only one of them is
operating.\footnote{More about the world described
in the following paragraphs can be found in
Marchildon (2004),
where it was introduced in terms of slightly
different experimental instruments.}

Consider a world which, as far as macroscopic
objects are concerned, is very similar to our own.
By this I mean that the laws of classical
mechanics, classical electrodynamics, and
thermodynamics apply to these objects with at
least as much generality as in our world.
They may even apply better, in the sense that their
scope may reach scales smaller than the
$10^{-9}$ to $10^{-10}$~m characteristic
of molecules and atoms in our world.  I shall not,
however, specify the microscopic structure of
the hypothetical world, except for one restriction
soon to be made.

In the hypothetical world, the macroscopic
objects sometimes behave in ways that cannot
be explained by the classical theories.  In
one class of situations, for instance, there are
devices (much like our Geiger counters) that click
when specific objects (like our radioactive
materials) are brought nearby.  We label the
former $D$ and the latter $E$.  Although the
clicks are random, their probability distributions
follow well-defined and reproducible laws.
We assume that these laws coincide with the
quantum-mechanical rules for the propagation
of wave packets.  If, for instance, there is
some material between $E$ and $D$, the number
and distribution of clicks are influenced just
like the quantum-mechanical theory of scattering
predicts. 

To account for these correlations, one can
envisage at least two very different explanatory
schemes.  In the first one, we postulate that
$E$ emits ``particles'' that are detected by $D$
after possibly interacting with intervening
objects.  In the second one, we postulate that
$D$ clicks in a way that is genuinely
fortuitous (Ulfbeck \& Bohr, 2001; Bohr,
Mottelson, \& Ulfbeck, 2004), the spatiotemporal
probability distribution of the clicks, however,
being dependent on the distribution of various
types of nearby macroscopic objects.

Now I make the assumption (and this is the
crucial way in which the hypothetical world differs
from our own) that the ``particles'' used in one
explanatory scheme to account for
the macroscopic correlations have no function
whatsoever in any attempt to explain the microscopic
structure of macroscopic objects.  That is, whether
matter is discrete or continuous at microscopic
scales, its small-scale constituents have nothing to do
with whatever is responsible for the clicks
described above.

How similar would the ether and the particles be,
as explanatory devices, in this hypothetical world?
Very much indeed.  Neither would have predictive
power not already contained in the alternative
explanations provided by the principle of relativity,
in one case, and probabilistic correlations, in
the other.  And both could be dispensed with
in a rational and completely articulated account
of nature.  Those who would keep the ether might do
so because of some prejudice in favour of absolute
simultaneity or motion.  Those who would keep the
particles might be influenced by the greater or
smaller number and types of them necessary to
explain the phenomena, or might find such contact
interactions more palatable.

Let us now leave the hypothetical world
and turn to the actual world, the one we live in.
Here quantum mechanics is also used to explain the
ultimate structure of macroscopic objects.
Moreover, it does so with the same mathematical
tools as the ones it uses to account for the
correlations described above.  That is, the state
spaces and parameters associated with the
``particles'' are also the ones (or at
least part of the ones) associated with the
building blocks of macroscopic objects.

In this context, what is the function of
Bohmian trajectories (or, for that matter, of
other interpretative schemes of quantum
mechanics)?  They provide us with one clear
way that the particles can behave so as to
reproduce the quantum-mechanical rules and,
therefore, the observable behaviour of macroscopic
objects.\footnote{Indeed from a Bohmian perspective,
the trajectories are just the kind of variables
that show up in a measurement, in sharp contrast
with the ether in electromagnetic theories.}
Although they could be dispensed with
in the hypothetical world, they cannot in the real
world unless, just like the ether was replaced by
the field, they are replaced by something that
can account for the structure of macroscopic
objects.

It has been argued that the emergence of the
notion of an autonomous field, connected with the
development of special relativity, has a parallel
in the emergence of the concept of information in
the context of quantum mechanics.  The motivation
for this is an important result recently obtained by
Clifton, Bub, and Halvorson (2003; Halvorson, 2004).
Working in the setting of $C^*$-algebras, these
investigators characterized the quantum theory by three
properties: (i) kinematic independence,
i.e.\ the commutativity of
the algebras of observables pertaining to distinct
physical systems; (ii) the noncommutativity of
an individual system's algebra of observables;
and (iii) nonlocality, i.e.\ the existence of entangled
states for spacelike-separated systems.  They
then showed that these properties are equivalent to
three information-theoretic constraints, namely,
the impossibility of superluminal information
transfer, of perfect broadcasting, and of
unconditionally secure bit commitment.

Drawing on this result, Bub (2005)
has proposed that quantum theory should be treated as
``\textit{a theory about the representation and
manipulation of information}'' (p.~557), where quantum
information is ``a new physical primitive not
reducible to the behaviour of mechanical systems
(the motion of particles and/or fields)'' (p.~546).
This, he argues, renders Bohmian trajectories
no more useful in quantum mechanics than the
ether is in special
relativity (Bub, 2004, p.~262):
\small
\begin{quote}
[J]ust as Einstein's analysis (based on the
assumption that we live in a world in which
natural processes are subject to certain
constraints specified by the principles of
special relativity) shows that we do not need
the mechanical structures in Lorentz's theory
(the aether, and the behaviour of electrons in
the aether) to explain electromagnetic phenomena,
so the [Clifton, Bub, and Halvorson] analysis
(based on the assumption that we live in a world
in which there are certain constraints on the
acquisition, representation, and communication
of information) shows that we do not need the
mechanical structures in Bohm's theory (the guiding
field, the behaviour of particles in the guiding
field) to explain quantum phenomena.
\end{quote}
\normalsize

To assess the validity of this claim, one
should point out that there is a
fundamental ontological difference between
field and information.  The electromagnetic
field, in the framework of special relativity,
is an autonomous entity that carries energy
and momentum.  Since Maxwell's equations have
solutions corresponding to vanishing charge
and current densities, the field can exist, in
principle, even in the complete absence of matter.
This is not the case with information, not in
the sense of Shannon at least.  To
exist, it needs some kind of material (or other)
support.  Whether in classical or quantum
mechanics, information is a functional on states of
objects.  It does not live autonomously.

This means that information-theoretic
considerations are relevant to the first
explanatory function of quantum mechanics,
the one that pertains to nonclassical
correlations of macroscopic objects.  But is
information, as ``a new physical primitive,''
of any help in the second explanatory
function, i.e.\ in accounting for the
structure of macroscopic objects?  It seems that
no proponent of the epistemic view would go so
far as suggesting that information is a fundamental
building block of nature, something objects are
made of.  This is very much unlike the
electromagnetic field, which at the turn of the
twentieth century was thought to account for part
or even for all the mass of charged particles
(McCormmach, 1970).  Hence the question about the
relevance of information to the second explanatory
function should be answered in the
negative.\footnote{Timpson
(2004) has provided an in-depth analysis of the Clifton,
Bub, and Halvorson result.  He first investigated
the extent to which the no bit-commitment constraint
is needed in characterizing quantum mechanics in the
framework of $C^*$-algebras.  He then examined
the relevance and generality of that formalism.
Closer to the aim of the present paper, he next
enquired whether viewing quantum mechanics as a
theory about the manipulation of information can
constitute an interpretation in an interesting sense.
Based on a distinction between the technical and
everyday senses of information, and
on the observation
that ``in both settings `information' functions as
an abstract noun, hence does not refer to a
particular or substance,'' his answer is largely
negative.}
%
%\newpage
\section{Discussion}
Several objections can be made to the claims that
Bohmian trajectories and the ether fulfill
distinct explanatory functions, and that information
is not a fundamental entity.  They must now be
addressed.

I have argued that quantum mechanics is used to
explain both (i) the nonclassical correlations
between macroscopic objects and (ii) the small-scale
structure of macroscopic objects.  But are these
two functions really different?  Suppose, for instance,
that we use the quantum theory to explain properties
of a macroscopic crystal, such as its elasticity or
heat capacity.  We should then make hypotheses on,
among other things, the atomic structure of the lattice
and the quantum-mechanical Hamiltonian.  But none of these
hypotheses can be tested directly.  They are tested
indirectly through their consequences on macroscopic
parameters such as elasticity constants or heat
capacity.  And the values of these parameters are
measured through correlations established between
experimental preparations and results displayed
by macroscopic pointers.

This is also the case with molecular properties.
Suppose that chemical analysis has revealed that
a given substance is chemically pure, so that we
ascribe its constitution to one type of molecule
only.  Properties of the substance, such as its
visible or infrared absorption, can then be explained
by applying the quantum theory to the electrons
and nuclei making up the molecule.  But again, none
of this can be tested directly.  Absorption
frequencies, in the end, show up as readings on some
macroscopic device, and the whole experimental
protocol reduces to correlations between macroscopic
preparations and macroscopic measurements.

I should readily admit that the empirical consequences of
both types of explanation provided by quantum
mechanics are of the same nature.  But the explanations
themselves are very different epistemologically,
as was illustrated in the last section by the
example of the hypothetical world.  The explanation
given of the structure of macroscopic objects, in
terms of atomic or subatomic constituents obeying
the laws of quantum mechanics, essentially
answers the question, What happens when objects are
repeatedly split?  That question seems
unavoidable in a complete understanding of
macroscopic objects.

In advocating the rejection of hidden variable
theories, Bub (2005, p.~557) argues that
\small
\begin{quote}
our measuring instruments \textit{ultimately
remain black boxes} at some level. That is,
a quantum description will have to introduce
a ``cut'' between what we take to be the ultimate
measuring instrument in a given measurement
process and the quantum phenomenon revealed by
the instrument, which means that the
measuring instrument is treated simply
as a probabilistic source of a range of labelled
events or ``outcomes''[.]
\end{quote}
\normalsize
As the phrase ``at some level'' indicates,
our measuring instruments are not total black boxes.
Indeed we can go a long way explaining the
properties of their parts on the basis of atomic
structure.  Should one argue that the atomic
structure is not to be taken literally,
he should be prepared to specify at what scale
ought the analysis of matter stop, or the
reality of objects dissolve.\footnote{This, by
the way, is related to the reason why the epistemic
view, \emph{even on its own terms}, won't solve the
measurement problem.  The epistemic view
is concerned with probabilities
of results of eventual macroscopic measurements.  But
it is not prepared to precisely specify what a
macroscopic apparatus is.  It won't tell us, for
instance, just how small an apparatus can be.  To
the credit of its proponents, none (as far as I know)
has proposed a purely arbitrary criterion like
``an apparatus must have a mass greater than
\mbox{1\,g}.''  But are there really any others?}

This brings us to a somewhat different objection
to the claims being made here.  What if the
structure of matter did not require explanation,
or at least could be accounted for by a very
different type of explanation than the one we
are used to?  It is well known that in the
history of science, criteria for what requires
explanation, or what counts as a valid explanation,
have often changed (Klein, 1972; Gardner, 1979; Cushing, 1990).
Gravitation through action
at a distance, considered impossible within the
seventeenth-century mechanistic worldview, became
less and less problematic in the eighteenth
century (McMullin, 1989).
Indeed the Laplacian school tried to account
for all terrestrial phenomena on the basis of
central forces which, though either attractive
or repulsive, were modeled on gravitation.
By then such forces were considered
mechanical (Fox, 1974).
In the late nineteenth century, mechanical
explanations were challenged both by
energetics (Ostwald, 1895)
and by the electromagnetic view of
matter (McCormmach, 1970).

Yet it seems that in all these instances,
answers were given to the question, What are
objects made of?  They could be made of point
particles acting on each other (partly at least)
without intermediaries.  Or else they could be made
of energy, or of electromagnetic fields.  But
again, information is not on a par with
such potential constituents.
Objects are not made of information.

Few people would go as far as advocating that
the small-scale structure of macroscopic objects
simply does not require explanation.  Yet something
close to this might be entailed by the idea of
\emph{genuine fortuitousness}.  The idea ``implies
that the basic event, a click in a counter, comes
without any cause and thus as a discontinuity in 
spacetime'' (Bohr, Mottelson, \& Ulfbeck, 2004, p.~405).
Indeed
\small
\begin{quote}
[i]t is a hallmark of the theory based on genuine
fortuitousness that it does not admit physical variables.
It is, therefore, of a novel kind that does not deal
with things (objects in space), or measurements, and
may be referred to as the theory of no
things.  (p.~410)
\end{quote}
\normalsize

Genuine fortuitousness, it turns out, could pretty
well fulfill the first explanatory function of
quantum mechanics, the one concerned with nonclassical
correlations of macroscopic objects.  Indeed it
would be quite unobjectionable, as an explanation
of these correlations, in the hypothetical world
I have described in Sec.~4.  But it fails to fulfill
the second explanatory function of quantum
mechanics.  When its proponents claim to eliminate
atoms or elementary particles, they seem always to have
in mind their alleged role in producing a click or
an ionisation track, rather than their role in
accounting for the structure of matter.  In fact
they cannot help contemplating the structure
of macroscopic counters, when for instance they point
out that ``the click involves such an immense number of
degrees of freedom that two clicks are never
identical'' (Ulfbeck and Bohr, 2001, p.~761).
One can immediately ask, How many degrees of freedom
are there?  What objects do they characterize?
Are these objects irreducible?  And so on.

To sum up, neither the ether nor Bohmian
trajectories have specific empirical
consequences.  Yet in addition to defining
a reference frame where simultaneity would
be absolute, the ether functioned as a kind
of support for electromagnetic phenomena.  That
role was transferred to the field when the ether
was discarded with the advent of special
relativity.
Bohmian trajectories, or other
interpretative schemes of quantum mechanics,
try to make the basic variables of the theory, in
terms of which the structure of macroscopic
objects is ultimately explained, intelligible.
This role, I have argued, cannot be dispensed with.
\section*{Acknowledgments}
It is a pleasure to thank Pierre Gravel, Karl Hess, and
Pierre Mathieu for comments and suggestions.  I am also
grateful to several anonymous referees whose comments
contributed in sharpening the ideas presented in this
paper.
%
%\newpage
\section*{References}
Bell, J. S. (1976).
How to teach special relativity.
Reprinted in J. S. Bell,
\textit{Speakable and unspeakable in quantum mechanics}
(pp.~67--80).
Cambridge: Cambridge University Press (1987).

\vspace{1ex}\noindent
Bennett, C. H., \& Brassard, G. (1984).
Quantum cryptography: public key distribution and coin tossing.
In \textit{Proceedings of the IEEE international conference
on computers, systems and signal processing} (pp.~175--179).
New York: IEEE.

\vspace{1ex}\noindent
Bohm, D. (1952).
A suggested interpretation of the quantum
theory in terms of `hidden' variables (I and II).
\textit{Physical Review, 85}, 166--193.

\vspace{1ex}\noindent
Bohm, D., \& Hiley, B. J. (1993).
\textit{The undivided universe}.
London: Routledge.

\vspace{1ex}\noindent
Bohr, A., Mottelson, B. R., \& Ulfbeck, O. (2004).
The principle underlying quantum mechanics.
\textit{Foundations of Physics, 34}, 405--417.

\vspace{1ex}\noindent
Brown, H. R. (2001).
The origins of length contraction: I. The
FitzGerald-Lorentz deformation hypothesis.
\textit{American Journal of Physics, 69}, 1044--1054.

\vspace{1ex}\noindent
Brown, H. R., \& Wallace. D. (2005).
Solving the measurement problem: de Broglie-Bohm
loses out to Everett.
\textit{Foundations of Physics, 35}, 517--540.

\vspace{1ex}\noindent
Bub, J. (2004).
Why the quantum?
\textit{Studies in History and Philosophy of Modern
Physics, 35}, 241--266.

\vspace{1ex}\noindent
Bub, J. (2005).
Quantum mechanics is about quantum information.
\textit{Foundations of Physics, 35}, 541--560.

\vspace{1ex}\noindent
Clifton, R., Bub, J., \& Halvorson, H. (2003).
Characterizing quantum theory in terms
of information-theoretic constraints.
\textit{Foundations of Physics, 33}, 1561--1591.

\vspace{1ex}\noindent
Cushing, J. T. (1990).
\textit{Theory construction and selection in modern
physics.  The S Matrix}.
Cambridge: Cambridge University Press.

\vspace{1ex}\noindent
Cushing, J. T. (1998).
\textit{Philosophical concepts in physics}.
Cambridge: Cambridge University Press.

\vspace{1ex}\noindent
Darrigol, O. (2000).
\textit{Electrodynamics from Amp\`{e}re to Einstein}.
Oxford: Oxford University Press.

\vspace{1ex}\noindent
DeWitt, B. S. (1970).
Quantum mechanics and reality.
\textit{Physics Today, 23}(9), 30--35.

\vspace{1ex}\noindent
Einstein, A. (1905).
On the electrodynamics of moving bodies.
Translated in \textit{The theory of relativity} (pp.~35--65).
New York: Dover (1952).

\vspace{1ex}\noindent
Einstein, A. (1949).
Autobiographical notes.
In P. A. Schilpp (Ed.),
\textit{Albert Einstein: philosopher-scientist} (pp.~1--95).
La Salle, IL: Open Court.

\vspace{1ex}\noindent
Everett, H. (1957).
`Relative state' formulation of quantum mechanics.
\textit{Reviews of Modern Physics, 29}, 454--462.

\vspace{1ex}\noindent
Fox, R. (1974).
The rise and fall of Laplacian physics.
\textit{Historical Studies in the Physical Sciences,
4}, 89--136.

\vspace{1ex}\noindent
Fuchs, C. A. (2002).
Quantum mechanics as quantum information
(and only a little more).
In A. Khrennikov (Ed.),
\textit{Quantum theory: reconsideration of foundations}
(pp.~463--543). 
V\"{a}xj\"{o}, Sweden: V\"{a}xj\"{o} University Press.
Also available as quant-ph/0205039.

\vspace{1ex}\noindent
Fuchs, C. A., \& Peres, A. (2000).
Quantum theory needs no `interpretation'.
\textit{Physics Today, 53}(3), 70--71.

\vspace{1ex}\noindent
Gardner, M. R. (1979).
Realism and instrumentalism in 19th-century atomism.
\textit{Philosophy of Science, 46}, 1--34.

\vspace{1ex}\noindent
Gr\"{u}nbaum, A. (1973).
\textit{Philosophical problems of space and time} (2nd ed.).
Dordrecht: Reidel.

\vspace{1ex}\noindent
Halvorson, H. (2004).
Remote preparation of arbitrary ensembles
and quantum bit commitment.
quant-ph/0310001.

\vspace{1ex}\noindent
Holland, P. R. (1993).
\textit{The quantum theory of motion}.
Cambridge: Cambridge University Press.

\vspace{1ex}\noindent
Klein, M. J. (1972).
Mechanical explanation at the end of the
nineteenth century.
\textit{Centaurus, 17}, 58--82.

\vspace{1ex}\noindent
Leggett, A. J. (2002).
Testing the limits of quantum mechanics:
motivation, state of play, prospects.
\textit{Journal of Physics: Condensed Matter, 14},
R415--R451.

\vspace{1ex}\noindent
Lorentz, H. A. (1909).
\textit{The theory of electrons}.
Mineola, NY: Dover (Reprint published 2003).

\vspace{1ex}\noindent
Marchildon, L. (2004).
Why should we interpret quantum mechanics?
\textit{Foundations of Physics, 34}, 1453--1466.

\vspace{1ex}\noindent
Maudlin, T. (1994).
\textit{Quantum non-locality and relativity}.
Oxford: Blackwell.

\vspace{1ex}\noindent
Maxwell, J. C. (1873).
\textit{A treatise on electricity and magnetism, Vols.~1 \& 2}.
New York: Dover (Reprint published 1954).

\vspace{1ex}\noindent
McCormmach, R. (1970).
H. A. Lorentz and the electromagnetic view of nature.
\textit{Isis, 61}, 459--497.

\vspace{1ex}\noindent
McMullin, E. (1989).
The explanation of distant action: historical notes.
In J. T. Cushing \& E. McMullin (Eds.),
\textit{Philosophical consequences of quantum theory}
(pp.~272--302).
Notre Dame: University of Notre Dame Press.

\vspace{1ex}\noindent
Ostwald, F. W. (1895).
Emancipation from scientific materialism.
Translated in M. J. Nye (Ed.),
\textit{The question of the atom} (pp.~337--354).
Los Angeles: Tomash/American Institute of Physics (1984).

\vspace{1ex}\noindent
Pais, A. (1982).
\textit{`Subtle is the Lord\ldots' The science and
the life of Albert Einstein}.
Oxford: Oxford University Press.

\vspace{1ex}\noindent
Paty, M. (1993).
\textit{Einstein philosophe}.
Paris: Presses universitaires de France.

\vspace{1ex}\noindent
Poincar\'{e}, H. (1898).
La mesure du temps.
Translated in \textit{The value of science} (Chap.~2).
New York: Dover (1958).

\vspace{1ex}\noindent
Poincar\'{e}, H. (1905).
Sur la dynamique de l'\'{e}lectron.
Reprinted in \textit{Oeuvres de Henri Poincar\'{e},
Vol.~9} (pp.~494--550).
Paris: Gauthier-Villars (1950--65). 

\vspace{1ex}\noindent
Rovelli, C. (1996).
Relational quantum mechanics.
\textit{International Journal of Theoretical Physics,
35}, 1637--1678.

\vspace{1ex}\noindent
Shor, P. W. (1994).
Algorithms for quantum computation: discrete
logarithms and factoring.
In \textit{Proceedings of the 35th annual symposium on
foundations of computer science} (pp.~124--134).
Los Alamitos, CA: IEEE. 

\vspace{1ex}\noindent
\mbox{'}t Hooft, G. (1999).
Quantum gravity as a dissipative deterministic system.
\textit{Classical and Quantum Gravity, 16}, 3263--79.

\vspace{1ex}\noindent
Timpson, C. G. (2004).
Quantum information theory and the foundations
of quantum mechanics.
Oxford University thesis.
quant-ph/0412063.

\vspace{1ex}\noindent
Tumulka, R. (2004).
Understanding Bohmian mechanics: a dialogue.
\textit{American Journal of Physics, 72}, 1220--1226.

\vspace{1ex}\noindent
Ulfbeck, O., \& Bohr, A. (2001).
Genuine fortuitousness.  Where did that click come from?
\textit{Foundations of Physics, 31}, 757--774.

\vspace{1ex}\noindent
Vandersypen, L. M. K., Steffen, M., Breyta, G., Yannoni, C. S.,
Sherwood, M. H., \& Chuang, I. L. (2001).
Experimental realization of Shor's quantum factoring
algorithm using nuclear magnetic resonance.
\textit{Nature, 414}, 883--887.

\vspace{1ex}\noindent
Van Fraassen, B. C. (1991).
\textit{Quantum mechanics: an empiricist view}.
Oxford: Oxford University Press.

\vspace{1ex}\noindent
Vermaas, P. E. (1999).
\textit{A philosopher's understanding of quantum
mechanics.  Possibilities and impossibilities
of a modal interpretation}.
Cambridge: Cambridge University Press.

\vspace{1ex}\noindent
Von Neumann, J. (1932).
\textit{Mathematical foundations of quantum mechanics}.
Princeton: Princeton University Press
(Translation published 1955).

\vspace{1ex}\noindent
Wallace, D. (2003).
Everett and structure.
\textit{Studies in History and Philosophy
of Modern Physics, 34}, 87--105.

\vspace{1ex}\noindent
Whittaker, E. T. (1951).
\textit{A history of the theories of aether and
electricity.  I. The classical theories}.
Los Angeles: Tomash/American Institute of Physics
(Reprint published 1987).

\vspace{1ex}\noindent
Zurek, W. H. (1991).
Decoherence and the transition from quantum to classical.
\textit{Physics Today, 44}(10), 36--44.
\end{document}